\documentclass[final,5p,times,twocolumn]{elsarticle}

\usepackage{graphics}
\usepackage{graphicx}
\usepackage{epsfig}
\usepackage{float}
\usepackage{xcolor}
\usepackage{amsmath}
\usepackage{lineno,hyperref}
\hypersetup{
	colorlinks   = true, 
	urlcolor     = red, 
	linkcolor    = blue, 
	citecolor   = blue 
}
\usepackage{amssymb}
\usepackage{multirow}
\usepackage{caption}
\include{defs}
\journal{Physica D}
\makeatletter
\makeatletter
\def\ps@pprintTitle{%
	\let\@oddhead\@empty
	\let\@evenhead\@empty
	\def\@oddfoot{\centerline{\thepage}}%
	\let\@evenfoot\@oddfoot}
\makeatother

\begin{document}

\begin{frontmatter}

\title{Identifying localized and spreading chaos in nonlinear disordered lattices by the Generalized Alignment Index (GALI) method}

\author[myprimaryaddress,mysecondaryaddress]{B.~Senyange}
\ead{b.senyange@muni.ac.ug}
\author[myprimaryaddress]{Ch.~Skokos\corref{mycorrespondingauthor}}
\ead{haris.skokos@uct.ac.za, haris.skokos@gmail.com}
\address[myprimaryaddress]{Nonlinear Dynamics and Chaos group, Department of Mathematics and Applied Mathematics, \\ University of Cape Town, Rondebosch, 7701 Cape Town, South Africa}
\address[mysecondaryaddress]{Department of Mathematics, Muni University, 725 Arua, Uganda}

\begin{abstract}
Implementing the Generalized Alignment Index (GALI) method of chaos detection we investigate the dynamical behavior of the nonlinear disordered  Klein-Gordon lattice chain in one spatial dimension. By performing extensive numerical simulations of single site and single mode initial excitations, for several disordered realizations and different disorder strengths, we determine the probability to observe chaotic behavior as the system is approaching its linear limit, i.e.~when its total energy, which plays the role of the system's nonlinearity strength, decreases. We find that the percentage of chaotic cases diminishes as the energy decreases leading to exclusively regular motion on multidimensional tori. We also discriminate between  localized and spreading chaos, with the former dominating the  dynamics for lower energy values. In addition, our results show that single mode excitations lead to more chaotic behaviors for larger energies compared to single site excitations. Furthermore, we demonstrate how the GALI method can be efficiently used to determine a characteristic chaoticity time scale for the system when strong enough nonlinearites lead to energy delocalization in both the so-called `weak' and `strong chaos' spreading regimes.      
\end{abstract}

\begin{keyword}
Disordered systems \sep Klein-Gordon system \sep Generalized Alignment Index (GALI) method \sep Chaotic motion \sep Localized chaos \sep Spreading chaos
\end{keyword}

\end{frontmatter}

\section{Introduction}
\label{sec:intro}

Anderson localization (AL) \cite{A58,KK93,EM08} is a general phenomenon appearing in linear disordered systems, leading to the exponential localization of the system's eigenmodes, and in turn to the halt of spreading of initially confined excitations, which has already been observed in numerous experimental set-ups \cite{WBLR97,CSG00,RZ03,GC05,SGAM06,BJZBHLCSBA08,HSPST08,KMZD11}. 

In the presence of nonlinearity the dynamics becomes more complicated as the system's normal modes (NMs) couple and chaos appears. Thus, the interplay of disorder and nonlinearity has attracted extensive attention in theory \cite{BW08,WZ09,F10,B11,CW11,MI12,I13,B14,MI14,MI15,C17,I17,CSZ21,CSD21}, numerical simulations \cite{S93,M98,KKFA08,PS08,FKS09,GS09,SKKF09,MAPS09,SF10,KF10,MP10,LBKSF10,JKA10,PF11,MAP11,BLSKF11,MAPS11,BLGKSF11,A11,ILF11,B12,LBF12,SGF13,MP13,M14,ABSF14,LIF14,TSL14,MKPM16,ATS16,ASBF17,SDRLM18,ATS18,SMS18,NTRSA19,MSS20,SPMS20} and experiments \cite{SBFS07,REFFFZMMI08,LAPSMCS08,LDTRZMLDIM11}. Two basic models have been considered in most of these numerical studies, the disordered Klein-Gordon (DKG) lattice of coupled anharmonic oscillators and the disordered discrete nonlinear Schr\"{o}dinger equation (DDNLS), with the latter being mainly used in theoretical studies.

One basic question which attracted extensive attention is related to the long-time, asymptotic fate of an initially localized excitation in a nonlinear disordered lattice. Most numerical studies \cite{S93,M98,PS08,FKS09,GS09,SKKF09,LBKSF10,BLSKF11} showed that nonlinearity eventually destroys AL, leading to subdiffusion spreading of the wave-packet, which is characterized by a power law increase of its second moment $m_2$ as $m_2 \propto t^{a_m}$, with $0 < a_m <1$. Nevertheless, indications of slowing down of such power laws, based on scaling analysis of numerical results, were reported in \cite{MAP11,MP13}. In addition, speculations about the eventual crossover of the dynamics to regular, quasiperiodic behavior were formulated in \cite{JKA10,A11}, while rigorous mathematical arguments about the  slowing down of spreading, or even of its complete halt, were presented in \cite{BW08,CSZ21}. 
More specifically, in \cite{BW08} it was analytically proven that the solutions of the DDNLS system remain localized for all times for weak enough nonlinearity, i.e.~when the system is very close to its linear limit. Furthermore, \cite{KKFA08} contains a proof that in the DDNLS model, at least part of the initially localized wave-packet remains localized and never spreads when the nonlinearity is strong enough to induce a nonlinear frequency shift of some oscillators' fundamental frequencies outside of the linear system's frequency band. This behavior was described as the `selftrapping' dynamical regime and has been numerically found not only for the DDNLS system but also for the DKG model \cite{FKS09, SKKF09,SF10,LBKSF10,BLSKF11}. In \cite{LBKSF10,F10,BLSKF11} two different spreading regimes were identified, namely the so-called `weak' and `strong chaos' regimes, respectively related to $a_m=1/3$ and $a_m=1/2$ for one-dimensional (1D) lattices  \cite{SF10,LBKSF10,F10}. Theoretical arguments leading to the derivation of these values were also presented in \cite{MI12,B14,MI14}. It is worth noting that in the case of 2D systems the weak and strong chaos regimes are respectively characterized by  $a_m=1/5$ and $a_m=1/3$ \cite{FKS09,LBF12,MSS20}. Despite the fact that the DDNLS model conserves two quantities (the total energy and the wave-packet's total norm), while the DKG system conserves only the system's total energy, both models exhibit the same dynamical regimes. It is worth noting that for small energies and oscillation amplitudes an approximate mapping between the two models exists \cite{KP92,K93,JR04,J06,LBKSF10}. 

Nowadays it is known that wave-packet spreading in nonlinear disordered lattices is a chaotic process resulting to the wave-packet's randomization and chaotization \cite{S93,MAPS09,B11,B12,SGF13,TSL14,SMS18,MSS20}. The computation of the maximum Lyapunov Characteristic Exponent (mLCE) \cite{BGGS80a,BGGS80b,S10} has been used to quantify the degree of chaoticity of disordered lattices \cite{PF11,B12,M14,TSL14,ATS18, NTRSA19}, while  a detailed analysis of the time evolution of the mLCE's estimator $\Lambda_1$ in \cite{SGF13,SMS18,MSS20},  showed power law decays $\Lambda_1 \propto t^{a_{\Lambda}}$, with $a_{\Lambda} \approx -0.25\, (-0.3)$ and $a_{\Lambda} \approx -0.37\, (-0.46)$ for the weak (strong) chaos case, respectively for 1D and 2D systems. These behaviors indicate the decline of the systems' chaoticity, which nevertheless dose not shows any signs of a potential crossover to regular dynamics (which is characterized by    $a_{\Lambda} \approx -1$), at least up to the computationally achieved (long) integration times. 

Another open issue is understanding the dynamical behavior of nonlinear disordered systems in the limit of weak nonlinearity, when the system is approaching its linear, integrable counterpart where chaos is not present  and AL appears. A basic question in this framework is if there exist initial conditions of the system which lead to chaotic behavior, and more importantly  wave-packet spreading, and if so, how their measure changes (decreases) as we approach the linear limit.

In \cite{MI12,MI14} a critical value of the nonlinearity strength in the DDNLS system was determined, below which excitations remained localized similarly to the linear case. Based on computations of the mLCE the probabilistic nature of finding random initial conditions which lead to chaotic behavior was investigated in \cite{PF11} for the DDNLS system, and in \cite{B12,M14} for the DKG and related models. There it was found that the percentage of chaotic orbits decreases and eventually vanishes when the system's nonlinearity strength diminishes. In the same spirit, an analysis was performed in \cite{ILF11}, which focused on the probability of obtaining AL below some small but finite nonlinearity level. That study was founded on a criterion for determining the delocalization of wave-packets based on computations of their participation number, whose effectiveness was criticized in \cite{B11}. 

In this work we try to also obtain a more global understanding of the localized and/or chaotic nature of initially confined excitations (more specifically we consider single site and single mode excitations) for  the 1D DKG model as the system's nonlinearity strength (i.e.~its total energy) diminishes, tending to zero. We focus our attention on the DKG model as it is computationally easier than the DDNLS system, which typically requires two orders of magnitude more computational time for reaching, with the same accuracy, the same final integration times. We distinguish between  localized and extended chaos, something which was not explicitly considered in previous publications, discriminating between energy excitations leading to regular behavior, localized chaos and to delocalized spreading chaotic wave-packets as was done for example in \cite{NTRSA19}.  

In order to determine the regular or chaotic nature of an orbit we rely on computations of the Generalized Alignment Index (GALI) method \cite{SBA07,SBA08,MSA12}, which is an efficient chaos indicator overcoming the problems of the estimation of the mLCE to clearly identify  chaotic behavior (e.g.~slow convergence to its limiting value; appropriate adjustments, based also on the final integration time, of its threshold value to point out chaos; power law decay of the index in the case of a spreading chaotic wave-packet instead of a saturation to a positive finite value). Furthermore, the localization or delocalization of a wave-packet is based on the inspection of its profile, as well as on the time evolution of its participation number $P$ and its second moment $m_2$. In addition, exploiting the advantages of the GALI method as an efficient chaos indicator, we implement it to quantify the decrease of chaos strength when wave-packets are evolving in the weak and strong chaos regimes.

The paper is organized as follows. In Sect.~\ref{sec:model} we describe the Hamiltonian of the DKG model, along with the main numerical techniques we use in our investigation, emphasizing the mLCE and the GALI methods. In Sect.~\ref{sec:small_en} we present numerical results about the behavior of the DKG model when the nonlinearity strength is decreased so that the system approaches its linear limit, both for single site and single mode excitations, while Sect.~\ref{sec:large_en} is devoted to the application of the GALI method to cases belonging to the weak and strong chaos regimes. Finally, in Sect.~\ref{sec:discussion} we summarize the findings of our work and discuss their significance. 

\section{Model and numerical techniques}
\label{sec:model}

\subsection{The Hamiltonian model}
\label{sec:Ham}

The 1D DKG lattice model of $N$ coupled anharmonic
oscillators is described by the Hamiltonian
\begin{equation}
	\label{eq:ham1}
	H({\boldsymbol{u}},{\boldsymbol{p}}) = \sum_{l=1}^N \left[ \frac{p_l^2}{2} +
	\frac{\epsilon_lu_l^2}{2} + \beta{u_l^4} +
	\frac{1}{2W}\left(u_{l+1} - u_l\right)^2 \right],
\end{equation}
where ${\boldsymbol{u}}= (u_1, u_2, \ldots, u_N)$ and ${\boldsymbol{p}}= (p_1, p_2, \ldots, p_N)$ are respectively the generalized positions and momenta. The fixed, random values of the coefficients $\epsilon_l$ are uniformly chosen  from the interval $\left[\frac{1}{2},\frac{3}{2}\right]$, determining the on-site potentials (a particular set of $\epsilon_l$, $l=1,2, \ldots, N$ is referred to as a disorder realization of the system), while $W$ and $\beta$ respectively quantify  the strength of disorder and  nonlinearity. In our study we consider only two values for $\beta$, namely $\beta=0$, for which the studied system becomes linear, and $\beta=1/4$, for the nonlinear version of the model studied in several publications in the past e.g.~\cite{FKS09,SKKF09,LBKSF10,BLSKF11,ILF11,SGF13,SMS18}, and impose  fixed boundary conditions, $u_0=u_{N+1}=p_0=p_{N+1}=0$.
The energy $h_l$ of oscillator $l$ is given by 
\begin{equation}
\label{eq:en_pa_site_1dkg}
	h_l=\frac{p_l^2}{2}+\frac{\epsilon_lu_l^2}{2}+
	\beta{u_l^4}+
	\frac{1}{4W}\left[ \left(u_{l-1}-u_l\right)^2+\left(u_{l+1}-u_l\right)^2 \right].
\end{equation} 

In this framework we typically follow the evolution of energy distributions created by the initial excitation of $L$  central oscillators at the same energy level $h_l=H/L$  by setting $p_l=\pm \sqrt{2H/L}$, with randomly assigned signs (for $L=1$ we always keep the $+$ sign) for the excited $L$ sites, and $u_l=0$ for all sites. Then, for the normalized energy distribution $h_l/H$ we compute its  second moment
\begin{equation}
\label{eq:m2}
	m_2=\sum_l(l-\bar{l} )^2\frac{h_l}{H}
\end{equation} 
with $\bar{l}=\sum_l\left(lh_l/H\right)$ denoting the centre of the distribution, which estimates the wave-packet's extent of spreading, and its participation number 
\begin{equation}
\label{eq:P}
	P=\frac{H^2}{\sum_lh_l^2},
\end{equation}
which quantifies  the number of highly excited sites. 

As the studied nonlinear version of \eqref{eq:ham1} has always $\beta=1/4$ the system's total energy  $H$ is used to regulate the nonlinearity strength of the system. Hamiltonian \eqref{eq:ham1} becomes linear when we neglect the nonlinear term $\beta{u_l^4}$ in \eqref{eq:ham1} (i.e.~taking $\beta=0$), while its nonlinear version with $\beta=1/4$ approaches the linear model  when the total energy $H$ tends to zero. Setting $u_l(t)=A_le^{-i\omega t}$, $l=1,2, \ldots, N$, and substituting it in the  equations of motion 
\begin{equation}
\label{eq:motionp}
	\frac{d{\boldsymbol{u}}}{dt} =  \frac{\partial H}{\partial{\boldsymbol{p}}},\qquad
	\frac{d{\boldsymbol{p}}}{dt} = - \frac{\partial H}{\partial{\boldsymbol{u}}},
\end{equation}
of the linear system ($\beta=0$), we get the linear eigenvalue problem 
\begin{equation}
\label{eq:eigval}
	\omega^2 A_l = \frac{1}{W}\left[- A_{l+1} + (W\epsilon_l+2)A_l - A_{l-1}\right].
\end{equation}
The system's NMs are the normalized eigenvectors $A_{\nu,l}$ (so that $\sum_lA_{\nu,l}^2=1$), $\nu=1,2,\ldots,N$,  and the frequencies of these  modes are the corresponding eigenvalues $\omega_{\nu}^2$.

\subsection{The maximum Lyapunov characteristic exponent}
\label{sec:LCE}

The set of Lyapunov Characteristic Exponents (LCEs), the so-called spectrum of LCEs, was introduced by Lyapunov \cite{L1892}, and became a fundamental tool for investigating the behavior of dynamical systems  \cite{O68,BGGS80a,BGGS80b,S10}. The spectrum of LCEs of an orbit in an autonomous Hamiltonian system of $N$ degrees of freedom consists of $2N$ values $\lambda_l$, $l=1,2,\ldots,2N$, which measure the mean exponential rate of growth (or shrinking) of small perturbations of the studied orbit. The LCEs come in pairs of values with opposite signs, $\lambda_l=-\lambda_{2N-l+1}$, and they are ordered as $\lambda_1 \geq \lambda_2 \geq \ldots \geq \lambda_{N-1} \geq \lambda_N = 0$ (see e.g.~\cite{S10} and references therein). If at least one LCE is positive, i.e.~$\lambda_1>0$, the orbit is characterized as chaotic, while if $\lambda_1=0$ (and consequently all the remaining LCEs are also zero) the orbit is regular.

The LCEs can be numerically obtained as time limits of appropriately computed quantities $\Lambda_i$, which are usually referred to as the finite time LCEs (ftLCEs), i.e.
\begin{equation}
\label{eq:ftMLCEs_limit}
	\lambda_l = \lim_{t \rightarrow \infty} \Lambda_l, \,\,\,\, l=1,2,\ldots, N,
\end{equation}
which can for example be evaluated by the so-called `standard method' \cite{BGGS80b,S10}. The interested reader can find in \cite{S10} more information on the practicalities related to the computation of the ftLCEs $\Lambda_i$, $i=1,2, \ldots, 2N$, along with pseudocodes for the actual estimation of $\lambda_i$.   In particular, the mLCE $\lambda_1$ is estimated as the limit for $t\rightarrow \infty$ of the finite-time mLCE (ftmLCE)
\begin{equation}
\label{eq:ftMLE}
	\Lambda_1 (t) = \frac{1}{t}\ln \frac{\lvert \lvert\boldsymbol{w}(t)
		\rvert \rvert}{\lvert \lvert\boldsymbol{w}(0) \rvert \rvert},
\end{equation} 
where $\boldsymbol{w} (t) = \delta \boldsymbol{z} (t) = \left( \delta \boldsymbol{u} (t),  \delta \boldsymbol{p} (t) \right)$ $= \left( \delta u_1 (t), \ldots, \delta u_N (t), \delta p_1 (t), \ldots, \delta p_N (t) \right)$ denotes the phase space perturbation vector from the orbit  $\boldsymbol{z} (t)=(\boldsymbol{u}(t),\boldsymbol{p}(t))$ at time $t$. In the case of regular orbits,  $\Lambda_1(t)$ tends to zero following the power law \cite{BGGS80b,S10}
\begin{equation}
\label{eq:LE_reg}
	\Lambda_1(t) \propto t^{-1},
\end{equation}
while for chaotic orbits, it tends to a non zero positive value.

The evolution of an initial deviation vector $\boldsymbol{w} (0)$ is governed by the so-called variational equations
\begin{equation}
\label{eq:var_Hess}
	\boldsymbol{\dot{w}} (t) = \begin{bmatrix}
		\dot{\delta u_l }(t) \\
		\dot{\delta p_l}(t)
	\end{bmatrix} =
	\begin{bmatrix}
		J_{2N} \boldsymbol{D}^2_{H} \left(\boldsymbol{z} (t) \right)
	\end{bmatrix} \cdot \boldsymbol{w} (t), \,\,\, l=1,2,\ldots, N,
\end{equation}
where $J_{2N} = \begin{bmatrix} 	0_N & I_N \\ 	-I_N & 0_N \end{bmatrix}$, with $I_N$ and  $0_N$ being respectively  the identity and the zero $N \times N$ matrices. Furthermore, $\boldsymbol{D}^2_{H} \left(\boldsymbol{z} (t) \right)$ is the $2N\times 2N$ Hessian matrix whose entries $\left[\boldsymbol{D}^2_{H} \left(\boldsymbol{z} (t) \right)\right]_{i, j} = \frac{\partial ^2 H}{\partial z_i \partial z_j } \bigg\rvert _{\boldsymbol{z}(t)}$ are evaluated at the position $\boldsymbol{z}(t)$ of the orbit in the system's phase space for all $i,j=1,2,\ldots, 2N$. The elements of  matrix $\begin{bmatrix}	J_{2N} \boldsymbol{D}^2_{H} \left(\boldsymbol{z} (t) \right) \end{bmatrix}$ in  \eqref{eq:var_Hess} depend on the evolution of the orbit $\boldsymbol{z}(t)$ but are independent of $\boldsymbol{w}(t)$. Therefore, the set of linear [with respect to $w_l(t)$]  equations \eqref{eq:var_Hess} have to be solved together with the system's Hamilton equations of motion \eqref{eq:motionp}. We note that in order to estimate  for example the first $k\leq 2N$ exponents, $k$ deviation vectors have to be integrated, although in our study we will focus only on the computation of the mLCE. 

\subsection{The generalized alignment index method}
\label{sec:GALI}

Although the computation of the mLCE is the most widely used technique for characterizing the regular or chaotic nature of orbits, its computational drawbacks, like for example its slow convergence to its limiting value, led to the development of a number of other, efficient chaos detection techniques, which make use of the solutions of the variational equations, like for example the fast Lyapunov indicator (FLI) and its variants \cite{FLG97,FGL97,B05,B06,LGF16,B16}, the mean exponential growth of nearby orbits (MEGNO) \cite{CS00,CHS031,CG16}, the relative Lyapunov indicator (RLI) \cite{SEE00,SESF04,SM16}, the smaller alignment index (SALI) \cite{S01,SABV03,SABV04} and its generalization, the GALI \cite{SBA07,SBA08,MSA12,SkMa16}.

In our study we will use the GALI method, which  was introduced in \cite{SBA07}, and proved to be a very efficient chaos detection technique as it has been successfully used for studying the chaoticity of several dynamical systems, see e.g.~\cite{BMC09,MA11,MR11,CPPSM17,MMS20,MGV21}. According to \cite{SBA07} the GALI of order $k$ (GALI$_k$), $2 \leq k \leq 2N$, is defined to be the volume of the generalized parallelogram having as edges $k$ normalized deviation vectors 
\begin{equation}
\label{eq:ini_dvd}
	\hat{\boldsymbol{w}}_{i}(t)=\frac{{\boldsymbol{ w}}_{i}(t)}{||{\boldsymbol{w}}_{i}(t)||}, \,\,\,\, i=1,2,\ldots, k,
\end{equation}
which are initially linearly independent. More specifically its value is computed as the norm of the wedge product of these vectors
\begin{equation}
\label{eq:gali}
	\mbox{GALI}_k(t)=||\hat{\boldsymbol{w}}_1(t)\wedge\hat{\boldsymbol{w}}_2(t)\wedge\cdots\wedge\hat{\boldsymbol{w}}_k(t)||,
\end{equation} 
where $||\cdot||$ denotes the usual vector norm. We note that the wedge product in \eqref{eq:gali} becomes zero and the GALI$_k$ vanishes when the deviation vectors  become linearly dependent. Practically the GALI$_k$ can be computed as the product of the singular values $v_i$, i.e.
\begin{equation}
\label{eq:gali_sv}
	\mbox{GALI}_k(t)=\prod_{i=1}^{k}v_i(t),
\end{equation}
of the $2N \times k$ matrix ${\bf W}(t)$ having as columns the coordinates of the $k$ normalized deviation vectors \eqref{eq:ini_dvd} \cite{SBA08}. Additional  practical information about the computation of the GALI$_k$ method can be found in \cite{GALIsite}. 

The behavior of the GALI$_k$ for regular orbits lying on $s$D  torus  ($s\leq N$) is given by \cite{SBA08}
\begin{equation}
\label{eq:gali_regular}
	\mbox{GALI}_k(t)\propto
	\begin{cases} 
	\mbox{constant} & {\textnormal {for}} \qquad 2 \leq k \leq s \\
	\frac{1}{t^{k-s}} & {\textnormal {for}} \qquad s < k \leq 2N-s \\
	\frac{1}{t^{2(k-N)}} & {\textnormal {for}} \qquad 2N-s < k \leq 2N, 
\end{cases}
\end{equation}
while for chaotic orbits GALI$_k$ goes to zero exponentially fast with an exponent depending on the first $k$ largest LCEs \cite{SBA07}
\begin{equation}
\label{eq:gali_chaos}
	\mbox{GALI}_k(t)
	\propto 
	\exp\left({-t\sum_{i=2}^k(\lambda_1-\lambda_i)}\right)
	\approx
	\exp\left({-t\sum_{i=2}^k(\Lambda_1(t)-\Lambda_i(t))}\right),
\end{equation}
where the values $\lambda_i$ of the MLEs are approximated by the related ftMLEs $\Lambda_i$.

In our simulations the equations of motion \eqref{eq:motionp} defining the evolution of an initial excitation (orbit) of the system, along with the variational equations \eqref{eq:var_Hess} for one or more initial perturbations (deviation vectors) are integrated together using the tangent map method \cite{SG10,GS11,GES12} and the two part split order four symplectic integrator ABA864 \cite{BCFLMM13}, which has already proved to be a very efficient and reliable method for multidimensional Hamiltonian systems \cite{SS18,DMMS19}. Typically, we perform numerical integrations up to a maximum time of $T = 10^9$ time units using lattices with  up to $N = 1\,000$ oscillators, taking care that the evolved wave-packet does not reach the lattice's boundaries. In all cases the integration time step $\tau$ is chosen so that the system's energy \eqref{eq:ham1} is conserved at an absolute relative energy error 
\begin{equation}
\label{eq:relative_error}
\left| \frac{H(t)-H(0)}{H(0)} \right| \lesssim 10^{-5},
\end{equation} 
for all $t$.

\begin{figure}[t]
	\centering
	\includegraphics[width=1.0\columnwidth,keepaspectratio]{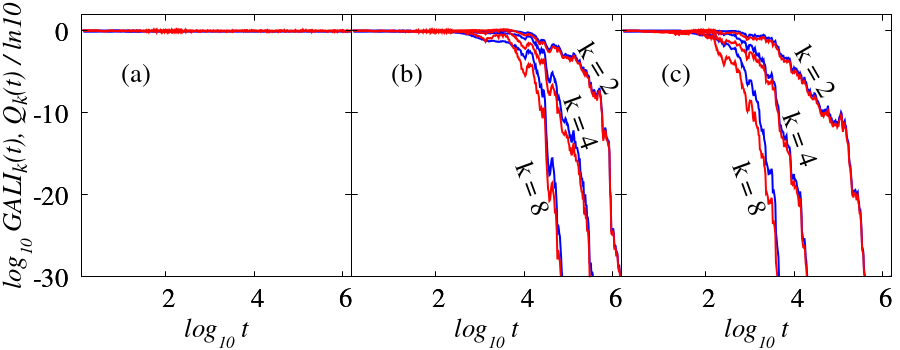}
	\caption{The time evolution of GALI$_k$ \eqref{eq:gali} (blue curves)  and $Q_k(t)=-t\sum_{i=2}^k(\Lambda_1(t)-\Lambda_i(t))$ (red curves) for $k=2, 4, 8$ of (a) a regular orbit of  system \eqref{eq:ham1} with  $W=3$, $\beta=0$, $L=37$, $H=3.7$ and two chaotic orbits respectively belonging to (b) the weak and (c) the strong chaos regime, with $W=3$, $\beta=1/4$, $L=37$, $H=0.37$ and $W=3$, $\beta=1/4$, $L=37$, $H=3.7$, when the same disorder realization is used in all cases. We note that in all panels the blue and red curves practically overlap. 
}
	\label{fig1:gali_behavior}
\end{figure}

In order to illustrate the behavior of the GALI$_k$ method, as well as its relation to the system's LCEs in the case of chaotic orbits \eqref{eq:gali_chaos}, we present in Fig.~\ref{fig1:gali_behavior} some representative computations of the index. More specifically, for the same disorder realization we consider evolutions for three particular cases of \eqref{eq:ham1}: a regular orbit for $W=3$, $\beta=0$, $L=37$, $H=3.7$ [Fig.~\ref{fig1:gali_behavior}(a)], an orbit belonging in the weak chaos spreading regime for  $W=3$, $\beta=1/4$, $L=37$, $H=0.37$ [Fig.~\ref{fig1:gali_behavior}(b)] and a chaotic orbit in the strong chaos regime for $W=3$, $\beta=1/4$, $L=37$, $H=3.7$ [Fig.~\ref{fig1:gali_behavior}(c)]. We note that the last two orbits respectively belong to the cases named $W1_{K}$ and $S2_{K}$ in \cite{SMS18}. In each panel of Fig.~\ref{fig1:gali_behavior} we plot the time evolution of GALI$_k$ (blue curves) for $k=2$, $4$ and $8$, along with the quantity $Q_k(t)=-t\sum_{i=2}^k(\Lambda_1(t)-\Lambda_i(t))$ (red curves) appearing in \eqref{eq:gali_chaos}.  The results confirm the validity of the relation GALI$_k(t) \propto \exp Q_k(t)$ (both for regular and chaotic orbits), and vividly demonstrate the ability of the index to efficiently discriminate between the two cases. In Fig.~\ref{fig1:gali_behavior}(a) all GALIs remain practically constant in agreement to the first equation of \eqref{eq:gali_regular}, indicating the regular nature of the orbit (something which is of course expected as in this case  system \eqref{eq:ham1} is linear, as $\beta=0$, and chaos is not present), while in both Figs.~\ref{fig1:gali_behavior}(b) and (c) GALIs eventually tend exponentially fast to zero (reaching extremely small values), denoting the chaotic nature of these orbits, in accordance to \eqref{eq:gali_chaos}. The higher chaoticity of the strong chaos case in Fig.~\ref{fig1:gali_behavior}(c) with respect to the weak chaos one in Fig.~\ref{fig1:gali_behavior}(b) becomes also evident by the faster decrease of the GALIs, which reach the same level of very small values faster than in Fig.~\ref{fig1:gali_behavior}(b). This behavior is also in agreement with the fact that the mLCE of the strong chaos cases reach higher values than the ones in the weak chaos regime (see Figs.~2(a) and 4(a) of \cite{SMS18}).

The results of Fig.~\ref{fig1:gali_behavior} clearly show that GALI$_2$, whose computation requires the numerical integration of only two deviation vectors, can efficiently determine the chaotic or regular nature of orbits.  We therefore only use GALI$_2$ in the  rest of this work  to study the chaotic behavior of system \eqref{eq:ham1}. In \cite{SBA07} it was shown that GALI$_2$ is practically equivalent to the SALI method  \cite{S01,SABV03,SABV04}, which  has  been successfully applied to study the chaotic behavior in  several dynamical systems (see for example \cite{SESS04,PBS04,BS06,CLMV07,MSCHJD07,SHC09,HW12,KKSK14,Z14}).

\section{Weak nonlinearity}
\label{sec:small_en}

In this section we implement the GALI$_2$  technique to study the  chaotic behavior of the nonlinear lattice system \eqref{eq:ham1} with $\beta=1/4$, for small nonlinearity strengths (i.e.~small $H$ values), by considering single site and single mode initial excitations.

\subsection{Single site excitations}
\label{sec:small_ss}

In order to investigate the dynamical behavior of the DKG system when its nonlinearity strength (quantified by the total energy $H$) decreases, so that the model approaches its linear counterpart, we perform numerical simulations by initially exciting one ($L=1$) central lattice site for different disorder realizations. In all considered cases we give to this site   energy $H$, setting also $\epsilon_l=1$ for this site in order to put it at the center of the interval $[1/2,3/2]$ from which all  other  $\epsilon_l$ values are randomly  chosen. We consider in our study several values of the energy $H$, ranging from  $H=0.3$ to $H=0.003$. In addition, trying to understand the possible influence of the disorder strength on these processes, we perform simulations both for  $W=4$ and $W=6$.

We classify a particular wave-packet evolution (i.e.~an orbit in the system's multidimensional phase space) as chaotic if the corresponding GALI$_2(t)$ value is practically zero [more specifically if GALI$_2(t)$ becomes $\leq 10^{-8}$] during the integration of the system up to the final considered time $T=10^9$, while, if this does not happen, the orbits is considered as non-chaotic/regular. One key aspect of our investigation is the differentiation between {\it localized chaos} and {\it extended or spreading chaos}. This distinction is based on the time evolution of   the participation number $P$ \eqref{eq:P} of the wave-packet. More specifically, a case is defined as spreading chaos when it is chaotic and   $P$  eventually does not remain constant, showing a clear tendency of progressive increase in time. On the other hand, the system shows localized chaos when GALI$_2$ classifies it as chaotic and at the same time, $P$  remains practically constant, fluctuating around some fixed value.

\subsubsection{Representative cases}
\label{sec:small_ss_rc}

\begin{figure}[t]
	\centering 
	\includegraphics[width=1.0\columnwidth,keepaspectratio]{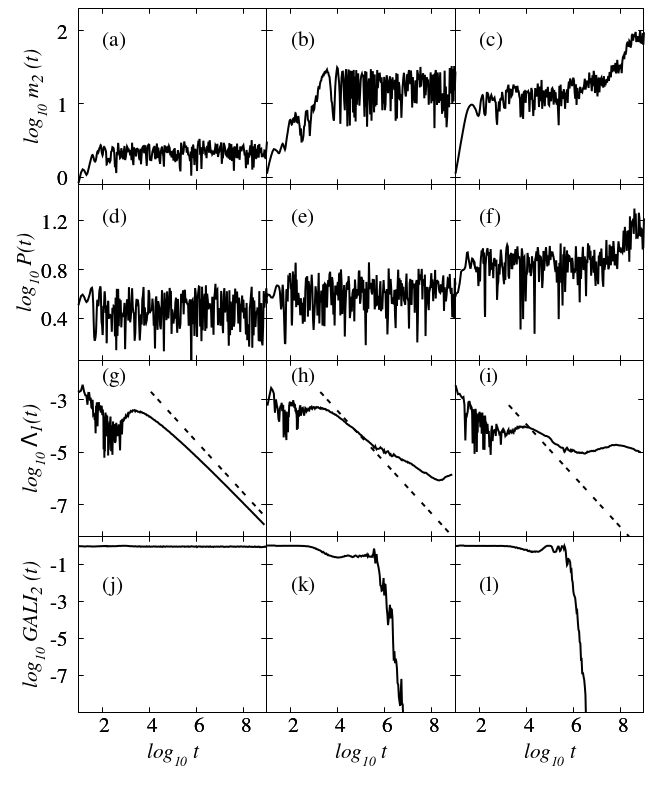}
	\caption{The time evolution of $m_2$ \eqref{eq:m2} [(a)-(c)], $P$ \eqref{eq:P} [(d)-(f)], $\Lambda_1$ \eqref{eq:ftMLE} [(g)-(i)] and GALI$_2$ \eqref{eq:gali} [(j)-(l)] for the same single site ($L=1$) excitation of the DKG system \eqref{eq:ham1} for $\beta=1/4$, $W=6$, $H=0.02$, for three different disorder realizations (one per column). The disorder realization of the left column [(a), (d), (g), (j)] leads to a regular evolution, while the one used in the middle [(b), (e), (h), (k)] and in the right [(c), (f), (i), (l)] column, respectively correspond to localized and spreading chaos. The straight dashed lines in (g), (h) and (i) guide the eye for slope $-1$, which characterizes regular dynamics. 	}
	\label{fig2:m2_P_L_gali}
\end{figure}

In Fig.~\ref{fig2:m2_P_L_gali} representative cases of these three dynamical behaviors, namely  regular dynamics [left column, panels (a), (d), (g), (j)], localized [middle column, panels (b), (e), (h), (k)] and spreading chaos [right column, panels (c), (f), (i), (l)] are presented for single site excitations of different disorder lattices, but for the same initial condition, total energy $H=0.02$ and disorder strength $W=6$, for the nonlinear ($\beta=1/4$) DKG system \eqref{eq:ham1}. For each case we follow the time evolution of the wave-packet's second moment  $m_2$ \eqref{eq:m2} [upper row, panels (a)-(c)], and participation number $P$ \eqref{eq:P} [second row, panels (d)-(f)] in order to describe the characteristics of the produced energy distribution profiles, along with the computation of the orbit's ftmLCE $\Lambda_1$ \eqref{eq:ftMLE} [third row, panels (g)-(i)] and GALI$_2$ \eqref{eq:gali} [lower row, panels (j)-(l)] for quantifying the system's chaoticity.

It is worth noting that, as we see from the results of Fig.~\ref{fig2:m2_P_L_gali}, different disorder realizations (but for the same set of parameters and initial conditions) lead to three different dynamical behaviors. In particular, the case presented in the left column of  Fig.~\ref{fig2:m2_P_L_gali} corresponds to a regular orbit. In this case the wave-packet remains localized for the duration of the evolution, as its measures of extent, $m_2$ [Fig.~\ref{fig2:m2_P_L_gali}(a)] and $P$ [Fig.~\ref{fig2:m2_P_L_gali}(d)] exhibit fluctuations around constant values. The regular nature of this orbit is clearly reflected in the evolution of the corresponding ftmLCE $\Lambda_1$ \eqref{eq:ftMLE} [Fig.~\ref{fig2:m2_P_L_gali}(g)], which constantly decreases to zero following a power law  $\Lambda_1 \propto t^{-1}$ [represented by the straight dashed line in all $\Lambda_1$ plots in Figs.~\ref{fig2:m2_P_L_gali}(g), (h) and (i)], appearing in the case of regular dynamics, as well as in its GALI$_2$ index [Fig.~\ref{fig2:m2_P_L_gali}(j)], which remains practically constant as is denoted in the first equation of \eqref{eq:gali_regular}.

Results for a simulation where we observe localized chaos is presented in panels (b), (e), (h) and (k) of Fig.~\ref{fig2:m2_P_L_gali}. In this case  the wave-packet's extent remains practically unchanged as  $m_2$ [Fig.~\ref{fig2:m2_P_L_gali}(b)] and $P$ [Fig.~\ref{fig2:m2_P_L_gali}(e)] do not show any signs of increase, while $\Lambda_1$ [Fig.~\ref{fig2:m2_P_L_gali}(h)] and GALI$_2$ [Fig.~\ref{fig2:m2_P_L_gali}(k)] show that the trajectory is chaotic. In particular, $\Lambda_1$ deviates from the $\Lambda_1 \propto t^{-1}$ decrease denoting regular behavior [dashed line in Fig.~\ref{fig2:m2_P_L_gali}(h)] when $\log_{10}t \gtrsim 6$. The chaotic nature of the orbit becomes evident, in an even more clear way, from the evolution of GALI$_2$ [Fig.~\ref{fig2:m2_P_L_gali}(k)], which abruptly decreases to zero for $\log_{10}t > 6$, justifying  in this way our decision to use this index in the following sections for efficiently identifying chaotic behavior.

In Figs.~\ref{fig2:m2_P_L_gali}(c), (f), (i) and (l) we present results for a disorder realization leading to an orbit exhibiting spreading chaos. As in the case of localized chaos of  Figs.~\ref{fig2:m2_P_L_gali}(b), (e), (h) and (k), the $\Lambda_1$ deviates from the $\Lambda_1 \propto t^{-1}$ decay [Fig.~\ref{fig2:m2_P_L_gali}(i)] and the GALI$_2$ eventually decreases abruptly to zero [Fig.~\ref{fig2:m2_P_L_gali}(l)], indicating the chaotic nature of the evolution. In addition, the spreading character of the wave-packet is reflected in the time evolution of both  $m_2$ [Fig.~\ref{fig2:m2_P_L_gali}(c)] and $P$ [Fig.~\ref{fig2:m2_P_L_gali}(f)], which, after a transient phase of bounded oscillations, show a clear increase towards the last stages of the simulation.

\begin{figure}[t]
	\centering
	\includegraphics[width=1.0\columnwidth,keepaspectratio]{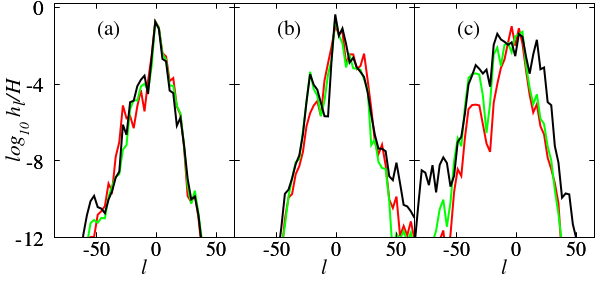}
	\caption{Normalized energy distributions $h_l/H$ at times 		$t = 10^5$ (red curves), $t = 10^7$ (green curves), and $t = 10^9$ (black curves) for the three cases of  of Fig.~\ref{fig2:m2_P_L_gali} leading to (a) regular evolution [left column of Fig.~\ref{fig2:m2_P_L_gali}], (b) localized chaos [middle column of Fig.~\ref{fig2:m2_P_L_gali}] and (c) spreading chaotic behavior [right column of Fig.~\ref{fig2:m2_P_L_gali}]. In all panels the initially excited oscillator at the center of the lattice is denoted by $l=0$. }
	\label{fig3:snapshots}
\end{figure}

In order to obtain a clearer understanding of the spatial evolution of the energy excitation in the three cases of Fig.~\ref{fig2:m2_P_L_gali}, we present in Fig.~\ref{fig3:snapshots} snapshots of the related normalized energy distributions, $h_l/H$, $l=1,2,\ldots, N$ (note a translation of the horizontal axis so that the initially excited site at the center of the lattice is located at $l=0$) at  times $t = 10^5$ (red curves), $t = 10^7$ (green curves), and $t = 10^9$ (black curves).

In Fig.~\ref{fig3:snapshots}(a) we see that for the regular case of Fig.~\ref{fig2:m2_P_L_gali} the three snapshots of the energy distributions practically overlap, having a pointy shape located at the position of the initial excitation. Thus, it becomes clear that the wave-packet's extent does not grow and always the same small number of sites are highly excited. This picture is in agreement with the computations of  the second moment [Fig.~\ref{fig2:m2_P_L_gali}(a)] and the participation number [Fig.~\ref{fig2:m2_P_L_gali}(d)] where both quantities attain small and practically constant values.  

From the results of Fig.~\ref{fig3:snapshots}(b) for the localized chaos case of Figs.~\ref{fig2:m2_P_L_gali}(b), (e), (h) and (k) we again see that the related energy distributions remain almost unchanged as time grows, as the energy profiles practically overlap, retaining again a pointy shape, which nevertheless is somewhat wider than the one in Fig.~\ref{fig3:snapshots}(a). This larger spatial extent of the localized wave-packet is clearly reflected on the larger, but bounded, values of $m_2$ observed in  Fig.~\ref{fig2:m2_P_L_gali}(b) with respect to Fig.~\ref{fig2:m2_P_L_gali}(a), as well as the similar behavior seen for the $P$ results depicted in Figs.~\ref{fig2:m2_P_L_gali}(e) and (d).

On the other hand, in the case of spreading chaos considered in Fig.~\ref{fig3:snapshots}(c), the energy distribution clearly grows in width when time increases, departing from its initial pointy shape  tending to form a more extended chapeau-like central region. This means that the number of highly excited sites increase in time leading to the increase of $P$ seen in Fig.~\ref{fig2:m2_P_L_gali}(f). The expansion of the wave-packet is also reflected in the growth of the related $m_2$ values in Fig.~\ref{fig2:m2_P_L_gali}(c).

\begin{figure}[t]
	\centering
	\includegraphics[width=1.0\columnwidth,keepaspectratio]{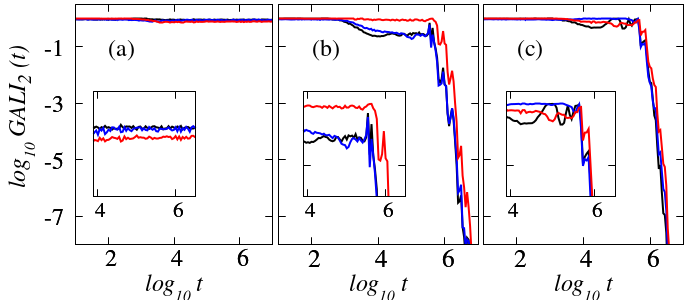}
	\caption{Time evolution of GALI$_2(t)$ for three different sets (black, blue and red curves) of random initial deviation vectors for (a) the regular case of Fig.~\ref{fig2:m2_P_L_gali}(j), (b) the localized chaos case of Fig.~\ref{fig2:m2_P_L_gali}(k), and (c) the spreading chaos case of Fig.~\ref{fig2:m2_P_L_gali}(l). The insets show a magnification of the main plots for the time interval $3.9 \leq \log_{10}t \leq 6.5$. 
	}
	\label{fig4:gali_indpdt}
\end{figure}

As a final, technical note we remark that the ability of the GALI$_2$ method to clearly and efficiently reveal the regular or chaotic nature of the wave-packet's evolution is practically independent of the set of the two orthonormal [so that GALI$_2(0)=1$] deviation vectors used for its computation. This is clearly seen in Fig.~\ref{fig4:gali_indpdt} where we plot the evolution of GALI$_2$ for three different sets  of random initial deviation vectors, in each one of the cases presented in Fig.~\ref{fig2:m2_P_L_gali} (results obtained by each set of vectors are plotted in different color: black, blue and red). In particular, in  Fig.~\ref{fig4:gali_indpdt}(a) we present results for the regular case of Fig.~\ref{fig2:m2_P_L_gali}(j), in  Fig.~\ref{fig4:gali_indpdt}(b) we see the evolution of GALI$_2$ for the localized chaos case of Fig.~\ref{fig2:m2_P_L_gali}(k), while in Fig.~\ref{fig4:gali_indpdt}(c) the spreading chaos case of  Fig.~\ref{fig2:m2_P_L_gali}(l) is considered. In each panel of Fig.~\ref{fig4:gali_indpdt} we see that the plotted curves either overlap or show a very similar trend. 

\subsubsection{Aggregate results}
\label{sec:small_ss_many}

Based on the analysis of the results of Fig.~\ref{fig2:m2_P_L_gali} we conduct a more general investigation of the dynamics of the DKG system \eqref{eq:ham1} when its energy $H$ decreases. For each considered energy value we follow the evolution of single site excitations for 100 different disorder realizations up to a final time $T=10^9$ time units. Using the corresponding evolution of GALI$_2(t)$ we determine the percentage $P_C$ of chaotic cases  resulting to GALI$_2(t)\leq 10^{-8}$ for $t\leq T$. Analyzing further these chaotic cases, by checking their $m_2$ and $P$ values, we discriminate between localized and spreading chaotic behavior (as we did for the cases presented in the middle and right columns of Fig.~\ref{fig2:m2_P_L_gali}) respectively determining their percentages $P_{CL}$ and $P_{CS}$, so that $P_C=P_{CL}+P_{CS}$.

\begin{figure}[t]
	\centering
	\includegraphics[width=0.495\columnwidth,keepaspectratio]{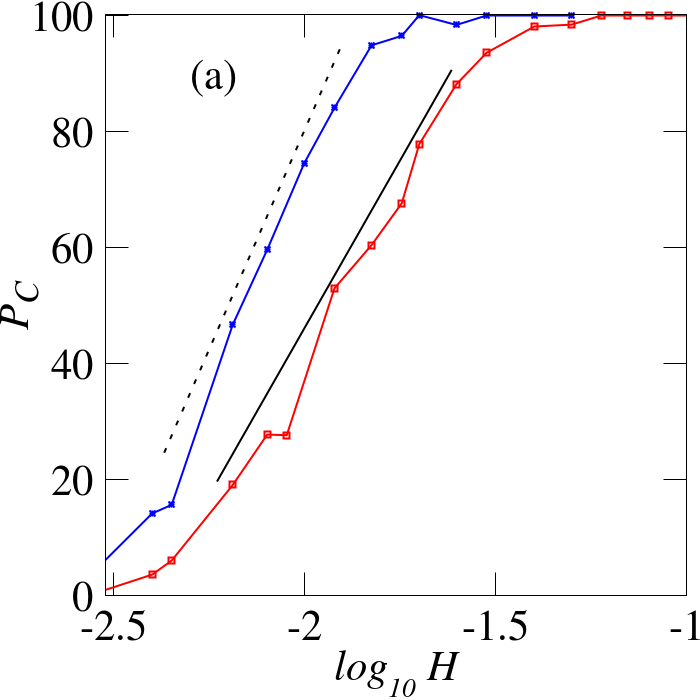}
	\includegraphics[width=0.495\columnwidth,keepaspectratio]{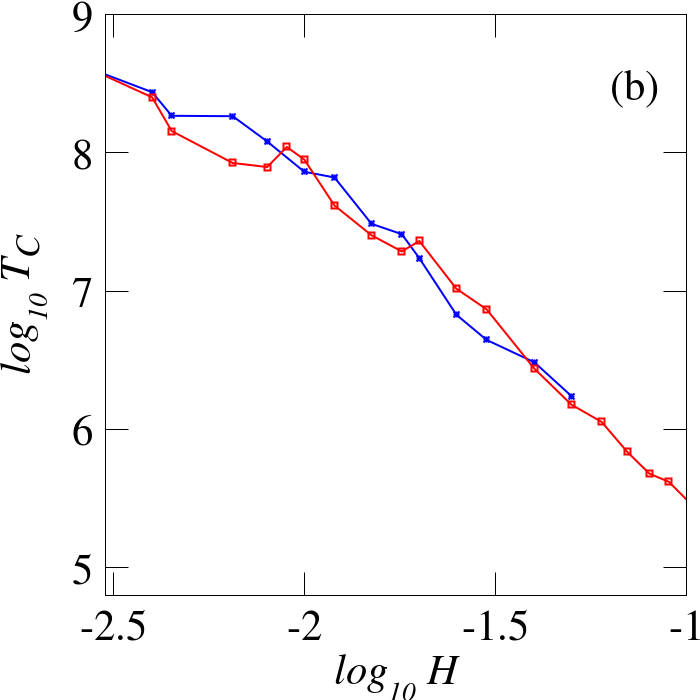}
	\caption{(a) Percentage $P_C$ of chaotic orbits for different energy values $H$ of the DKG model \eqref{eq:ham1} obtained by single site excitations of 100 different disorder realizations for  disorder strengths $W=4$ (blue points/curve) and $W=6$ (red points/curve). The continuous (dashed) black line indicates slope 120 (150). (b) The average [over the chaotic cases of (a)] time $T_C$ needed for GALI$_2$ to become $\leq 10^{-8}$ vs.~$H$ for $W=4$ (blue points/curve) and $W=6$ (red points/curve). In both panels data points are line connected in order to facilitate the visualization of the underlying trends.
	}
	\label{fig5:prop_chaos}
\end{figure}

In Fig.~\ref{fig5:prop_chaos}(a) we see the dependence of $P_C$ on the energy $H$ of the DKG system \eqref{eq:ham1} for disorder strengths $W=4$ (blue points/curve) and $W=6$ (red points/curve). For both cases $P_C$ increases with increasing  nonlinearity strength and it becomes $P_C=100\%$ at a high energy whose value depends on the disorder strength, i.e.~it is larger for $W=6$. On the other hand, approaching the linear limit of the DKG system by decreasing the energy, we observe a decrease of the percentage of chaotic orbits, which equivalently means that more disorder realizations accommodate regular behavior. This decrease is more abrupt for $W=4$, described by a slope close to $150$ [dashed black line in Fig.~\ref{fig5:prop_chaos}(a)], than in the $W=6$ case, for which the decrease is approximated by a slope $\approx 120$ [continuous black line in Fig.~\ref{fig5:prop_chaos}(a)]. A clear tendency of $P_C$ to become zero for small $H$ values is evident for both cases, implying that the system's phase space is almost completely occupied by invariant tori leading to quasiperiodic, regular motion. These findings are in agreement to the empirical and numerical arguments provided in \cite{JKA10} for the infinite random DDNLS model.

It is worth noting that the blue curve ($W=4$) is always above the red one ($W=6$) when $P_C \neq 100\%$, indicating that for the same nonlinearity strength (energy $H$) more chaos is present for $W=4$. This behavior can be understood in the following way. A single site excitation results to the excitation of several NMs, whose nonlinear interaction is responsible for the chaotic behavior of the wave-packet. As the spatial extent of the, nevertheless localized, NMs increases when $W$ decreases \cite{KK93,KF10,SPMS20}, more NMs are excited by single site excitations for $W=4$ than for $W=6$. On top of that, the wider extent of these NMs lead to stronger interactions between them and in turn, to higher level of chaos as is observed in Fig.~\ref{fig5:prop_chaos}(a). For the same reasons, higher energies are needed in the $W=6$ case to reach the fully chaotic level of $P_C=100\%$, as more localized NMs (with respect to the $W=4$ case) need a stronger nonlinear interaction to lead to well defined chaotic evolution.        

In Fig.~\ref{fig5:prop_chaos}(b) we present the dependence on $H$ of the average time $T_C$ needed for the chaotic cases of Fig.~\ref{fig5:prop_chaos}(a)  to clearly demonstrate their chaotic nature, i.e.~the time for which GALI$_2(t)$ becomes $\leq 10^{-8}$. We see that for both $W=4$ and $W=6$, $T_C$ shows a similar behavior: it increases as $H$ decreases approaching $T_C=10^9$ (the final integration time of our simulations) for very small $H$ values. This, implies that, although the number of cases exhibiting chaos increases as $W$ decreases, the time need for this chaos to be observed does not seem to be influenced by $W$, at least of course for the two values of $W$ considered here. Definitely this point deserves further investigation, something which we plan to address in a future publication by performing a similar analysis for more $W$ values.   

\begin{figure}[t]
	\centering
	\includegraphics[width=0.495\columnwidth,keepaspectratio]{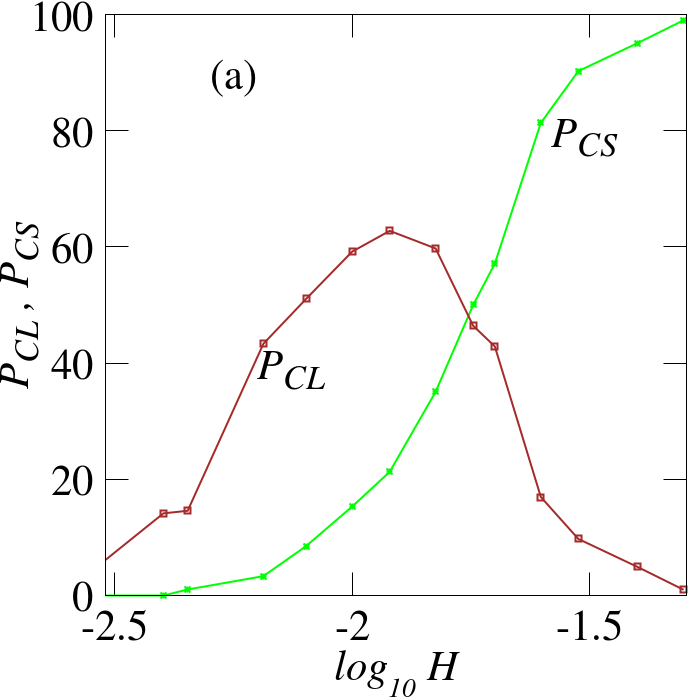}
	\includegraphics[width=0.495\columnwidth,keepaspectratio]{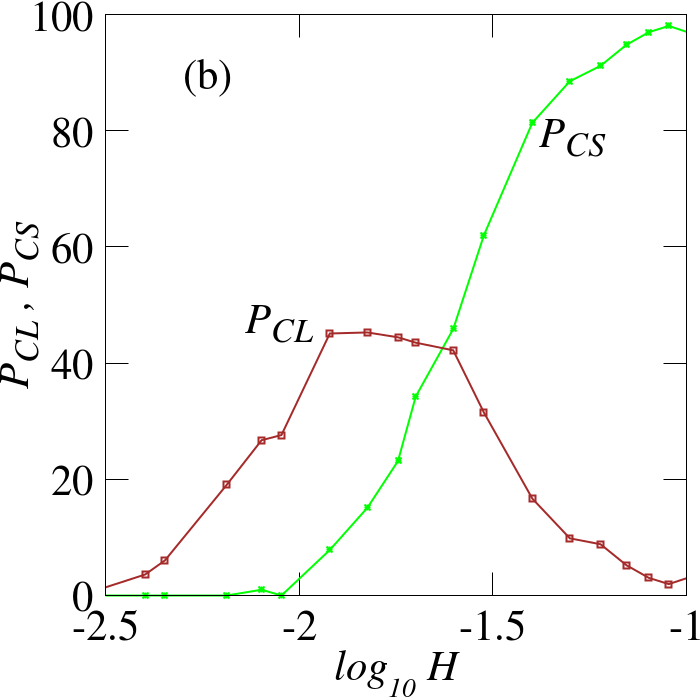}
	\caption{The percentage of cases of  Fig.~\ref{fig5:prop_chaos}(a) exhibiting localized chaos ($P_{CL}$, brown points/curves) and spreading chaos ($P_{CS}$, green points/curves) vs.~$H$ for (a) $W=4$ and (b) $W=6$. In both panels data points are line connected in order to facilitate the visualization of the underlying trends. }
	\label{fig6:prop_chaos}
\end{figure}

Looking closer to the cases leading to chaos, we distinguish between localized and spreading chaos by respectively plotting in Fig.~\ref{fig6:prop_chaos} their percentages, $P_{CL}$ (brown points/curves) and $P_{CS}$ (green points/curves), as a function of the system's energy $H$ for $W=4$ [Fig.~\ref{fig6:prop_chaos}(a)] and $W=6$ [Fig.~\ref{fig6:prop_chaos}(b)]. For both $W$ values, localized chaos dominates the dynamics for small energies, but as the system's nonlinearity grows the fraction $P_{CS}$ of spreading chaos cases increases, eventually reaching   $P_{CS}=100\%$, which means that for strong enough nonlinearities all cases exhibit chaotic spreading ($P_C=P_{CS}=100\%$).

By comparing the two panels of Fig.~\ref{fig6:prop_chaos}, we see that for $W=4$ the portion of localized chaos cases reaches higher values compared to the one observed for $W=6$. This means that whenever chaotic behavior appears for $W=6$, it predominately corresponds to spreading chaos. Possibly this happens because the system's NMs are more localized than in the $W=4$ case and so, when their interactions become powerful enough to introduce chaos, they are sufficiently strong to also lead to energy spreading.

\subsection{Single mode excitations}
\label{sec:small_sm}

In order to investigate further the chaotic behavior of the DKG system \eqref{eq:ham1}, we complement our study by also presenting results for single mode initial excitations. To do so, we order the NMs in space by increasing value of their center-of-norm coordinate $\bar{l}=\sum_l l A_{\nu,l}^2$, where $A_{\nu,l}$ is the amplitude of NM $\nu$, with $\nu =1,2, \ldots, N$. The NM's spatial extent has been reported to be maximized for eigenfrequencies positioned in the middle of the $\omega_{\nu}^2$ spectrum \cite{KK93,HSPST08,KF10}. We therefore initially excite a NM whose center is located around the middle of the lattice and whose eigenfrequency $\omega_{\nu}^2$ lies within the middle one third of the spectrum $[1/2,3/2+4/W]$, by attributing to it the total  energy $H$ of the system, while all other modes are not exited at all.

\begin{figure}[t]
	\centering 
	\includegraphics[width=1.0\columnwidth,keepaspectratio]{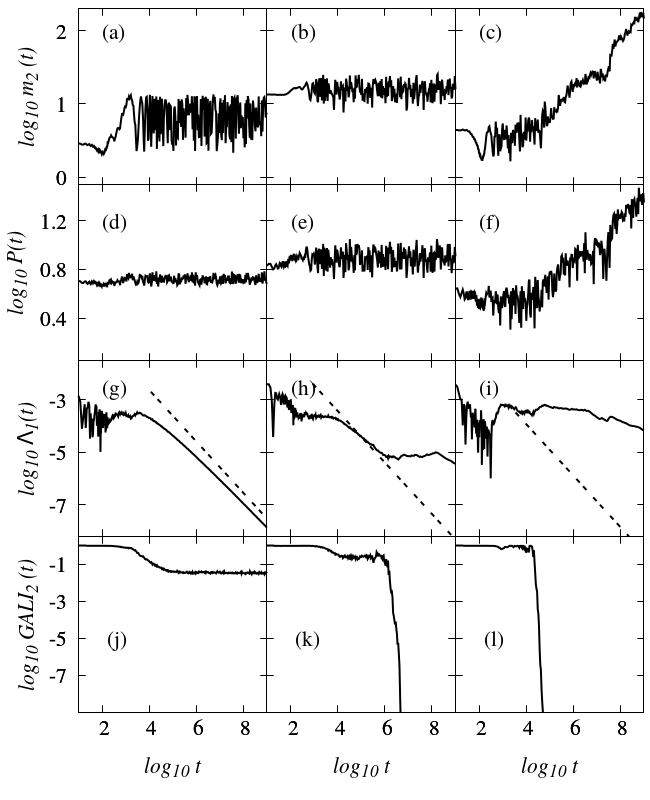}
	\caption{Similar to Fig.~\ref{fig2:m2_P_L_gali} but for single mode excitations of the DKG system \eqref{eq:ham1} for $\beta=1/4$, $W=6$, $H=0.1$, for three different disorder realizations (one per column).
 	}
	\label{fig27:m2_P_L_gali_sm}
\end{figure}

\begin{figure}[t]
	\centering
	\includegraphics[width=1.0\columnwidth,keepaspectratio]{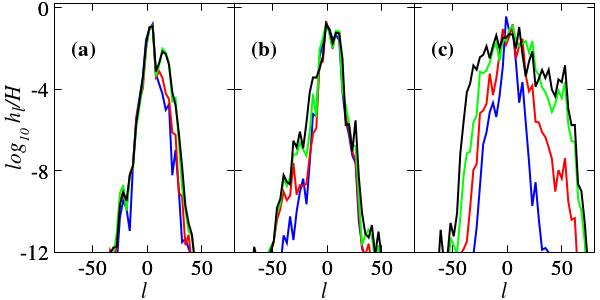}
	\caption{Normalized energy distributions $h_l/H$ at times $t = 0$ (blue curves), $t = 10^7$ (red curves), $t = 10^8$ (green curves), and $t = 10^9$ (black curves) for the three cases of  of Fig.~\ref{fig27:m2_P_L_gali_sm} leading to (a) regular behavior [left column of Fig.~\ref{fig27:m2_P_L_gali_sm}], (b) localized chaos [middle column of Fig.~\ref{fig27:m2_P_L_gali_sm}] and (c) spreading chaotic behavior [right column of Fig.~\ref{fig27:m2_P_L_gali_sm}]. }
	\label{fig38:snapshots_sm}
\end{figure}

As in the case of single site excitations (Sect.~\ref{sec:small_ss_rc}), we can find single mode excitations leading to regular or chaotic (localized or spreading) behaviors. In Fig.~\ref{fig27:m2_P_L_gali_sm}  we present (in a similar way to Fig.~\ref{fig2:m2_P_L_gali}) some representative cases of single mode excitations, for three different disorder realizations, resulting to regular dynamics [left column, panels (a), (d), (g), (j)], as well as to localized [middle column, panels (b), (e), (h), (k)] and spreading chaos [right column, panels (c), (f), (i), (l)]. In all cases Hamiltonian \eqref{eq:ham1} has $\beta=1/4$, $W=6$, $H=0.1$ and $N=1,000$, while the center $\bar{l}$ and the participation number $P$ of the initially excited mode along with its eigenvalue $\omega^2_{\nu}$ are $\bar{l}\approx 503.5$, $P \approx 5.1$, $\omega^2_{\nu} \approx 1.285$ [left column of Fig.~\ref{fig27:m2_P_L_gali_sm}], $\bar{l}\approx 502.4$, $P \approx 7.4$, $\omega^2_{\nu} \approx 1.458$ [middle column of Fig.~\ref{fig27:m2_P_L_gali_sm}] and $\bar{l}\approx 500.6$, $P \approx 4.6$,  $\omega^2_{\nu} \approx 1.318$ [right column of Fig.~\ref{fig27:m2_P_L_gali_sm}]. The corresponding energy distributions are seen in Fig.~\ref{fig38:snapshots_sm} for $t = 0$ (blue curves), $t = 10^7$ (red curves), $t = 10^8$ (green curves), and $t = 10^9$ (black curves). For the single mode excitation leading to regular dynamics we see that its energy profile does not practically change in time [Fig.~\ref{fig38:snapshots_sm}(a)] and consequently $m_2$ [Fig.~\ref{fig27:m2_P_L_gali_sm}(a)] and $P$ [Fig.~\ref{fig27:m2_P_L_gali_sm}(d)] mildly oscillate around constant values, while both $\Lambda_1$ [Fig.~\ref{fig27:m2_P_L_gali_sm}(g)] and GALI$_2$ [Fig.~\ref{fig27:m2_P_L_gali_sm}(j)] denote the regular nature of the orbit. The energy profiles of the localized chaos case [Fig.~\ref{fig38:snapshots_sm}(b)]  do not also practically change in time and  $m_2$ [Fig.~\ref{fig27:m2_P_L_gali_sm}(b)] and $P$ [Fig.~\ref{fig27:m2_P_L_gali_sm}(e)] fluctuate again around constant values, but now, due to the motion's chaotic nature,   $\Lambda_1$ [Fig.~\ref{fig27:m2_P_L_gali_sm}(h)] eventually deviates from the $\Lambda_1 \propto t^{-1}$ law [dashed line in Fig.~\ref{fig27:m2_P_L_gali_sm}(h)] and GALI$_2$ [Fig.~\ref{fig27:m2_P_L_gali_sm}(k)] vanishes for $\log_{10} t \gtrsim 7$. Finally, for the presented case of spreading chaos we see a clear increase in the width of its energy distribution [Fig.~\ref{fig38:snapshots_sm}(c)] resulting to the growth of the  $m_2$ [Fig.~\ref{fig27:m2_P_L_gali_sm}(c)] and $P$ [Fig.~\ref{fig27:m2_P_L_gali_sm}(f)] values, accompanied by a significant deviation of the ftmLCE from the  $\Lambda_1 \propto t^{-1}$ law [Fig.~\ref{fig27:m2_P_L_gali_sm}(i)] and a rather fast and abrupt decrease of  GALI$_2$ [Fig.~\ref{fig27:m2_P_L_gali_sm}(l)].

\begin{figure}[t]
	\centering
	\includegraphics[width=0.495\columnwidth,keepaspectratio]{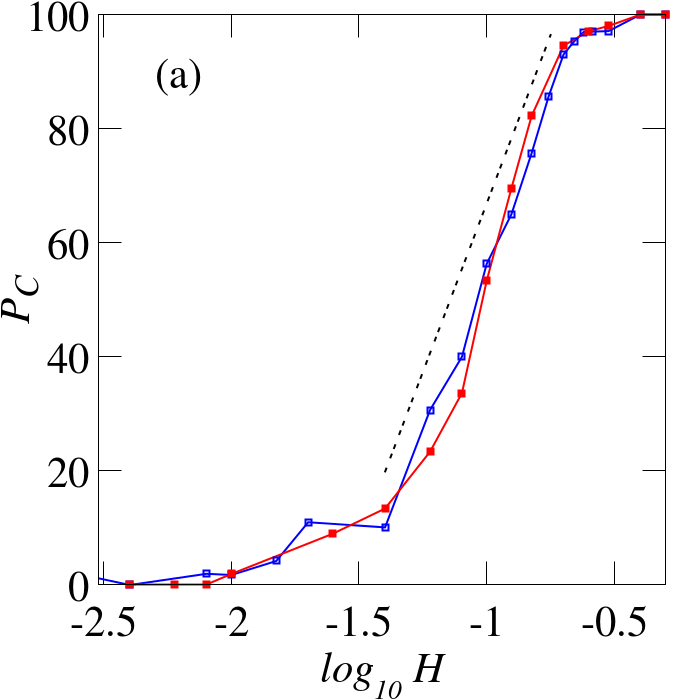}
	\includegraphics[width=0.495\columnwidth,keepaspectratio]{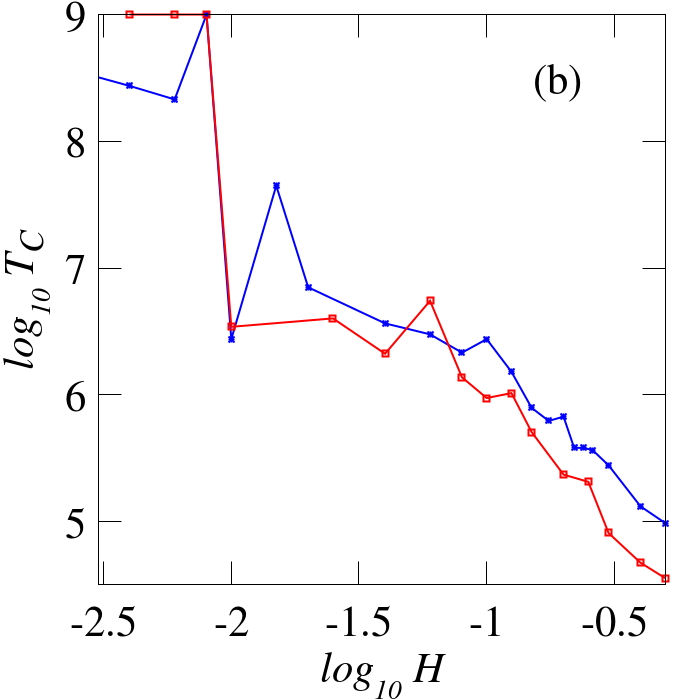}
	\caption{Similar to Fig.~\ref{fig5:prop_chaos} but for single mode initial excitations. The dotted black line in (a) indicates slope $120$. 
	}
	\label{fig7:prop_chaos}
\end{figure}

We also perform a similar analysis to the one presented in Sect.~\ref{sec:small_ss_many} for single site excitations, by obtaining statistical results over 100 different disorder realizations. In Fig.~\ref{fig7:prop_chaos}(a) we present results for the dependence of the fraction of chaotic cases, $P_C$, on the system's energy $H$ for disorder strengths $W=4$ (blue points/curve) and $W=6$ (red points/curve). Similarly to the case of single site excitations in  Fig.~\ref{fig5:prop_chaos}(a), $P_C=100\%$ for high energy values, and shows a  decrease as $H$ diminishes. This decrease is  characterized in  Fig.~\ref{fig7:prop_chaos}(a) by a $120$ slope (dotted black line) in its larger extent, although a change to a slower decrease is observed for small $H$ values ($\log_{10}H \lesssim -1.5$). 

Comparing Figs.~\ref{fig5:prop_chaos}(a) and \ref{fig7:prop_chaos}(a) we see that for small enough nonlinearities, so that not all simulations lead to chaos, single mode excitations  result to less chaotic behavior with respect to single site excitations for the same $H$ value, as in the latter case  more than one NMs are excited from the beginning. Thus, in the case of single site excitations mode-mode interactions are present from the beginning of the evolution leading to stronger coupling between modes in the presence of the same nonlineartity (despite the localized nature of the NMs), which in turn leads to strong chaotic interactions and wave-packet delocalization. This behavior becomes more pronounced when the NMs' extend increase (i.e.~for smaller $W$ values). On the other hand, in the case of single mode excitations, chaos is again introduced by the nonlinear interaction and overlap of NMs, which is in general weaker because initially only one mode is  excited.  This leads to smaller percentages of chaotic cases with respect to single site excitations for the same energy value [for example, from  Figs.~\ref{fig5:prop_chaos}(a) and \ref{fig7:prop_chaos}(a) we see that for $\log_{10} H \approx -1$ all single site excitations practically lead to chaos ($P_C\approx 100 \%$), while for single mode excitations we observe a weaker chaotic behavior as $P_C \approx 55\%$],  which also does not seem to depend on the considered values of $W$ as the results in Fig.~\ref{fig7:prop_chaos}(a) for $W=4$ and $W=6$ practically overlap. 

The average time $T_C$ required for the chaotic cases to reveal their nature, i.e.~the time needed for GALI$_2$ to become less that $10^{-8}$, shows a well defined increase as $H$ decreases [Fig.~\ref{fig7:prop_chaos}(b)], which does not seem to depend drastically on the value of $W$, similarly to what was observed for  single site excitations [Fig.~\ref{fig5:prop_chaos}(b)].

\begin{figure}[t]
	\centering
	\includegraphics[width=0.495\columnwidth,keepaspectratio]{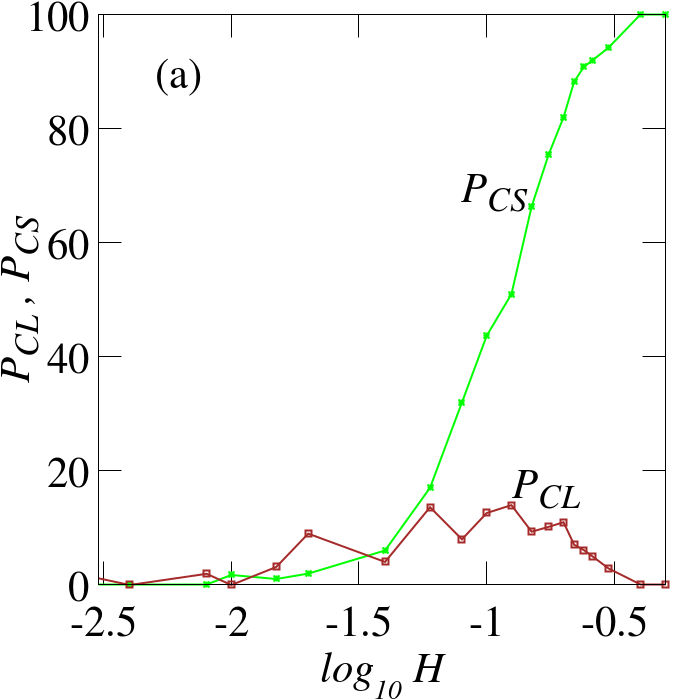}
	\includegraphics[width=0.495\columnwidth,keepaspectratio]{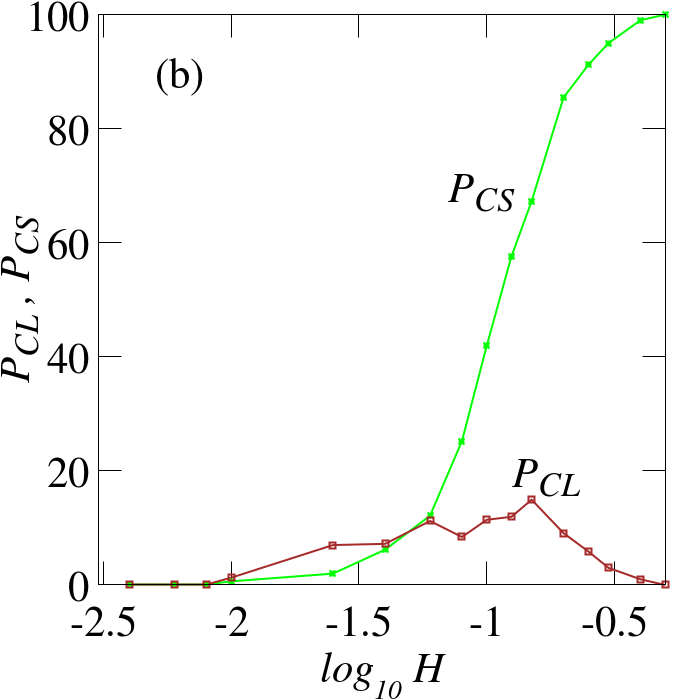}
	\caption{Similar to Fig.~\ref{fig6:prop_chaos} but for single mode initial excitations. }
	\label{fig8:prop_chaos}
\end{figure}

Fig.~\ref{fig8:prop_chaos} shows the percentages $P_{CL}$ of localized (brown points/curves) and $P_{CS}$ of spreading chaos (green  points/curves) for disorder strengths $W=4$ [Fig.~\ref{fig8:prop_chaos}(a)] and $W=6$ [Fig.~\ref{fig8:prop_chaos}(b)]. Unlike the case of single site excitations (Fig.~\ref{fig6:prop_chaos}), orbits leading to spreading chaos dominate the portion of chaotic cases compared to those where chaos is localized. Furthermore, not large differences are observed between the $W=4$ and $W=6$ cases. Again, at very high energies ($\log_{10}H \gtrsim -0.5$)  practically all chaotic orbits of the system exhibit spreading chaos.

From the comparison of Figs.~\ref{fig6:prop_chaos} and \ref{fig8:prop_chaos} it is worth noting that when chaotic behaviors start appearing as $H$ increases, they lead initially to localized chaos, as the nonlinearity strength should increase considerably in order to permit intense nonlinear interactions, which eventually lead to energy spreading and delocalization. Furthermore, we see that single site excitations leading to chaos appear for smaller energies when $W$ is smaller (i.e.~larger localization length of NMs permits more mode-mode interactions in the presence of nonlinearity), and we also observe higher percentages of  localized chaos  for $W=4$ (brown curves in Fig.~\ref{fig6:prop_chaos}). On the other hand, in the case of single mode excitations (Fig.~\ref{fig8:prop_chaos}) we again see a slightly higher percentage of localized chaos for small $H$ values, but eventually the dynamics is characterized by spreading chaotic behaviors for $\log_{10}H \gtrsim -1.2$, as  the energy becomes high enough to induce strong nonlinear effects.

\section{Weak and strong chaos spreading regimes}
\label{sec:large_en}

In the previous section, the GALI$_2$ method proved to be a very efficient and reliable tool for investigating the changes in the dynamical behavior of the DKG model as the system approaches its linear limit. Here we will also use this index to study the characteristics of chaos for strong enough nonlinearities in the weak and strong chaos spreading regime \cite{LBKSF10,F10,BLSKF11}. 

In \cite{SGF13,SMS18} it was shown that, as an initially localized wave-packet spreads, in both the weak and the strong chaos regimes, it becomes less chaotic without showing any signs of a crossover to regular dynamics (at least up to the largest time scales reached by the performed numerical simulations).  This behavior was established by the time evolution of the system's ftmLCE \eqref{eq:ftMLE}, which showed a power law decrease, $\Lambda_1 \propto t^{a_{\Lambda}}$ with $a_{\Lambda} \approx -0.25$ ($a_{\Lambda} \approx -0.3$) for the weak (strong) chaos regime. The fact that $a_{\Lambda}$ attains constant values, which are different from the $a_{\Lambda} =- 1$ observed in the case of regular motion (see e.g.~\cite{S10}), clearly indicates that the dynamics remain always chaotic. A characteristic time scale quantifying the time needed for the system to show its chaotic nature is obtained through the so-called `Lyapunov time' $T_{\Lambda}=\Lambda_1^{-1}$ (see e.g.~\cite{S10}). For the weak and strong chaos cases the Lyapunov time is 
\begin{equation}
\label{eq:L_t_DKG}
	T_{\Lambda}(t)=\frac{1}{\Lambda_1(t)}\propto \frac{1}{t^{a_{\Lambda}}}=
	\begin{cases} 
	t^{~0.25}, & {\textnormal{for weak chaos,}}\,\,\,\\
		t^{~0.3}, & {\textnormal{for strong chaos.}} 
		\end{cases}
\end{equation}
This equation clearly denotes that the system becomes less chaotic as time grows since $T_{\Lambda}$ grows.

We can also use GALI$_2$ in order to define a characteristic time scale of chaos, by following the procedure developed in \cite{MBS13}. To do so, we start the GALI$_2$ computation by considering two random orthonormal deviation vectors  $\hat{\boldsymbol{w}}_1(0)$, $\hat{\boldsymbol{w}}_2(0)$ so that GALI$_2(0)=1$. Then, whenever GALI$_2(t)$ becomes less than a very small value (in our case we set this value to be GALI$_2= 10^{-8}$) we reinitialize its computation by substituting the current deviation vectors $\hat{\boldsymbol{w}}_{1}(t)$ and $\hat{\boldsymbol{w}}_{2}(t)$ with the initially used pair  $\hat{\boldsymbol{w}}_1(0)$, $\hat{\boldsymbol{w}}_2(0)$, setting in this way  GALI$_2(t) = 1$. Thereafter, we let these vectors develop under the current  dynamics and follow again the evolution of GALI$_2$.

\begin{figure}[t]
	\centering
	\includegraphics[width=1.0\columnwidth,keepaspectratio]{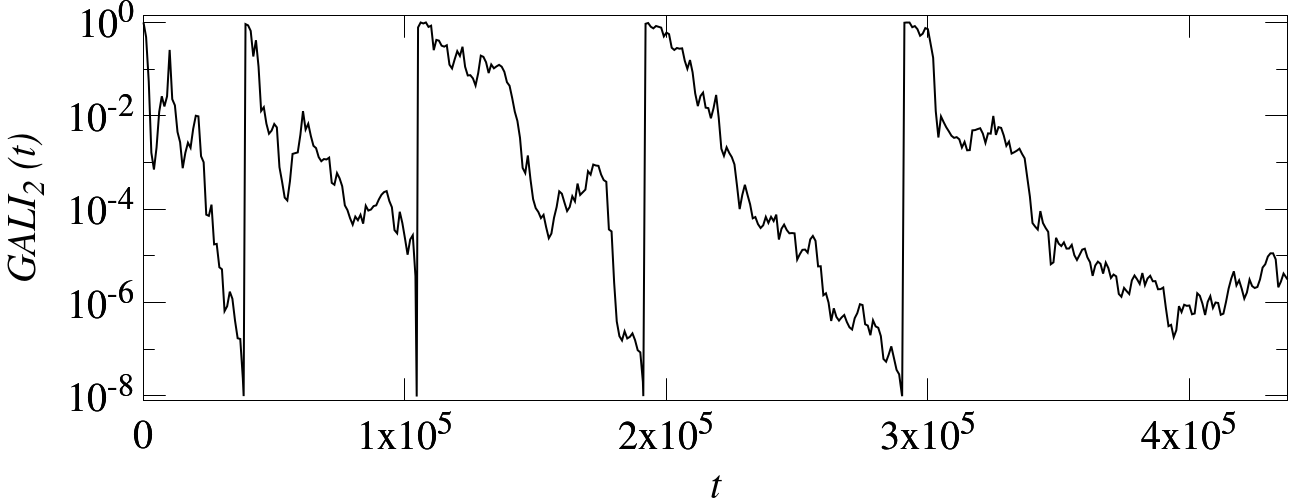}
	\caption{Time evolution of the reinitialized GALI$_2(t)$ of a representative orbit in the strong chaos regime of the DKG model \eqref{eq:ham1} with $W=3$, $L=37$, $\beta=1/4$ and $H=3.7$. 
	}
	\label{fig9:repeated_sali}
\end{figure}

In Fig.~\ref{fig9:repeated_sali} we present the time evolution of the reinitialized GALI$_2(t)$ for a representative case with parameters $W=3$, $L=37$, $\beta=1/4$ and $H=3.7$  belonging in the strong chaos regime. As we have already explained chaotic behavior is characterised by the exponential decay of GALI$_2$ [see Eq.~\eqref{eq:gali_chaos}]. Therefore, the time $T_G$ needed for the  GALI$_2$ to become $\leq 10^{-8}$ can be considered as an indicator of the system's chaoticity strength. From the results of Fig.~\ref{fig9:repeated_sali} it becomes evident that $T_G$ grows as time increases (the time difference between successive reaches of GALI$_2$ at GALI$_2=10^{-8}$ augments), a clear indication that the dynamics becomes less chaotic. 

In \cite{MBS13} this procedure was implemented in order to identify regular and chaotic epochs in the evolution of orbits in a time dependent Hamiltonian system, as in such systems the nature of orbits can be altered during their evolution. On the other hand, in autonomous Hamiltonian systems these alternations cannot happen as and orbit is either regular or chaotic (maybe weakly chaotic or sticky, as is called sometimes, but nevertheless chaotic). So what is the point of implementing this GALI$_2$ reinitialization procedure in the DKG model which is an autonomous system? Although Hamiltonian \eqref{eq:ham1} is an autonomous model of $N$ degrees of freedom the initial excitation of only $L$ ($L\ll N$) sites at the center of the lattice, as well as the subsequent energy spreading, which extends to more and more sites (with decreasing energy per excited site as the total energy remains fixed while the number of excited sites increases in time) without reaching the boundaries of the lattice (i.e.~without exciting all the $N$ degrees of freedom), means that in practice we have a conservative Hamiltonian system with an increasing number of active degrees of freedom. For this reason we  encounter  cases of chaotic motion whose chaoticity decrease  as time grows. Such cases are described by a power law decrease of the related ftmLCE $\Lambda_1$ \eqref{eq:ftMLE}, which is different from the one  ($\Lambda_1 \propto t^{-1}$) observed in the case of regular motion. Thus, the process depicted in Fig.~\ref{fig9:repeated_sali}  can be used for achieving an alternative way of quantifying the decrease of the system's chaoticity, through the  capture of the potential changes  in the reinitialization time $T_G$. A decreasing $T_G(t)$ would imply that the system becomes more chaotic, while an increasing $T_G(t)$, as is the case in  Fig.~\ref{fig9:repeated_sali}, signifies that the system becomes less chaotic.

\begin{figure}[t]
	\centering
	\includegraphics[width=0.495\columnwidth,keepaspectratio]{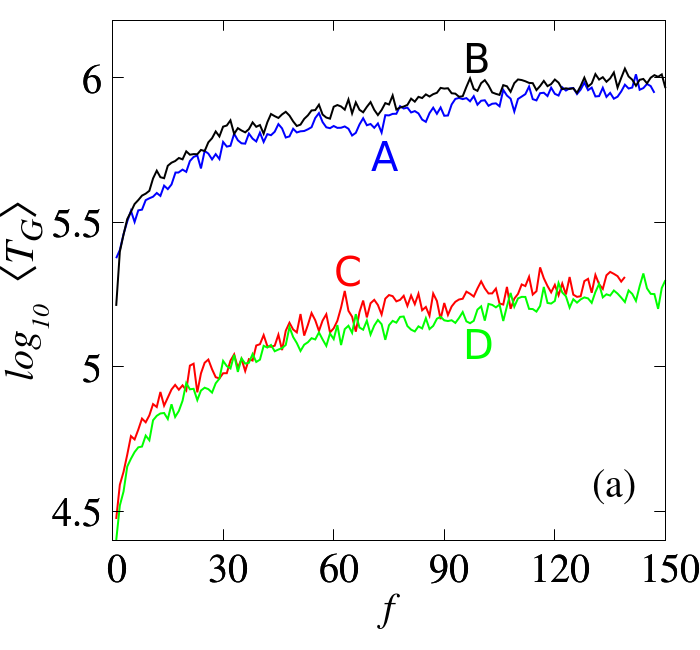}
	\includegraphics[width=0.495\columnwidth,keepaspectratio]{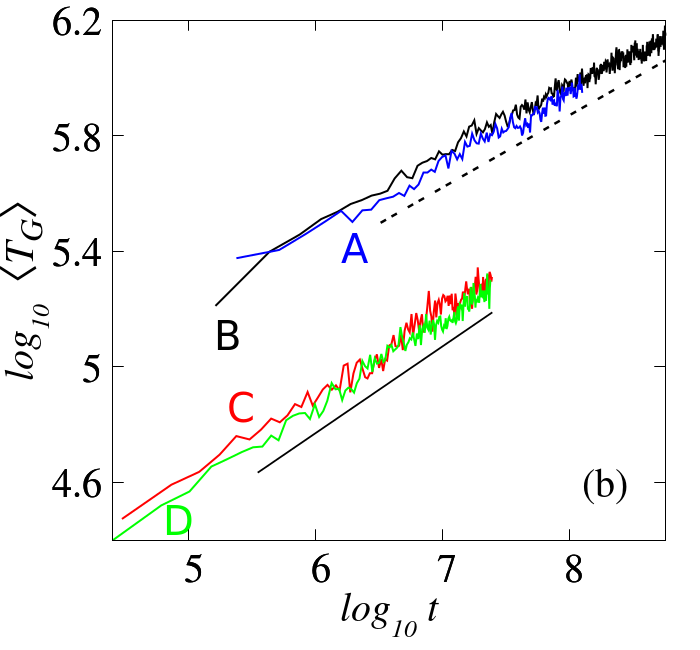}
	\caption{Average  (over 100 disorder realizations) reinitialization time $T_G$ plotted against (a) the number $f$ of reinitializations of  GALI$_2$ and (b) the integration time $t$ of the DKG system \eqref{eq:ham1} for the weak chaos cases {\bf A} (blue curves) and {\bf B} (black curves), as well as the strong chaos cases {\bf C} (red curves) and {\bf D} (green curves) (see text for more details). The straight black lines in (b) guide the eye for slopes $0.25$ (dashed line) and $0.3$ (solid line).  
	}
	\label{fig10:av_delay_t}
\end{figure}

In Fig.~\ref{fig10:av_delay_t} we present average (over 100 disorder realizations) results ($\langle T_G \rangle$) for the evolution of the GALI$_2$ reinitialization time $T_G$ for two weak chaos and two strong chaos cases. More specifically we consider the weak chaos cases of Hamiltonian \eqref{eq:ham1} with  $W=3$, $\beta=1/4$, $L=37$, $H=0.37$ (case  {\bf A}) and  $W=4$, $\beta=1/4$, $L=1$, $H=0.4$ (case  {\bf B}), as well as the strong chaos cases  $W=2$, $\beta=1/4$, $L=83$, $H=8.3$ (case  {\bf C}), and   $W=3$, $\beta=1/4$, $L=37$, $H=3.7$ (case  {\bf D}). We note that all these cases have already been presented in \cite{SMS18} where the were named $W1_{K}$ (case  {\bf A}), $W2_{K}$ (case  {\bf B}), $S1_{K}$ (case  {\bf C}) and $S2_{K}$ (case  {\bf D}). 

In Fig.~\ref{fig10:av_delay_t}(a) we plot the average value $\langle T_G \rangle$ as a function of the number $f$ of reinitializations of GALI$_2$, while in Fig.~\ref{fig10:av_delay_t}(b) $\langle T_G \rangle$ is plotted with respect to the actual time $t$. Results for the two weak chaos cases and the two strong chaos arrangements are similar, indicating that the behavior of $\langle T_G \rangle$ does not depend on the individual cases but it is related to the particular dynamical regime. The results of Fig.~\ref{fig10:av_delay_t} show a clear distinction between the weak and strong chaos cases, as in the former $\langle T_G \rangle$ attain much higher values, indicating that the system is less chaotic, as it requires more time to show clear signs of chaoticity. This difference is also an indirect validation of the appropriateness of the terms `weak' and `strong chaos regimes' used to describe the two dynamical behaviors. In Fig.~\ref{fig10:av_delay_t}(a) we see that, as the number $f$ of GALI$_2$ reinitializations grow, $\langle T_G \rangle$ exhibits a slow, but consistent increase indicating that the chaoticity of the system decreases with time. Furthermore, $\langle T_G \rangle$ remains finite for the entire duration of the simulations, showing that the dynamics of the system does not pass to regular behavior.

The results of Fig.~\ref{fig10:av_delay_t}(b) clearly suggest that the growth of $\langle T_G \rangle$ in time can be described by a power law of the form $\langle T_G \rangle \propto t^{a_G}$. Actually fitting with straight lines the results of Fig.~\ref{fig10:av_delay_t}(b) we  get $a_G=0.246 \pm 0.005$ (Case {\bf A}), $a_G=0.250  \pm 0.001$ (Case {\bf B}), $a_G=0.298 \pm 0.006$ (Case {\bf C}) and $a_G=0.299 \pm 0.007$ (Case {\bf D}). It is worth noting that the power law dependencies of $T_{\Lambda} (t)$ \eqref{eq:L_t_DKG} approximate quite well the time evolution of $\langle T_G \rangle$, both for the weak [dashed line in Fig.~\ref{fig10:av_delay_t}(b)] and the strong chaos case [continuous line in Fig.~\ref{fig10:av_delay_t}(b)], indicating that $\langle T_G \rangle(t) \propto T_{\Lambda} (t)$, and suggesting that the GALI$_2$ reinitialization time can be used to define a characteristic chaoticity time scale for the DKG system.

\section{Summary and conclusions}
\label{sec:discussion}

We investigated the chaotic behavior of the 1D DKG  model \eqref{eq:ham1} for various values of its total energy $H$ (which plays the role of the nonlinearity strength parameter) and its disorder strength $W$. We performed extensive numerical simulations of the propagation of single site and single mode excitations and obtained average results over  100 disorder realizations in each parameter setting. In our investigations we implemented the GALI$_2$ \eqref{eq:gali} chaos detection method to efficiently and accurately determine the chaotic nature of the produced wave-packets, while the evaluation of their second moment $m_2$ \eqref{eq:m2} and participation number $P$ \eqref{eq:P} helped us  specify the localized or expanding behavior of energy excitations. 

Based on these computations we were able to establish the probabilistic nature \cite{ILF11,PF11,B12,M14} of the appearance of chaotic or regular behaviors when the system's nonlinearity decreases, leading the DKG model closer to its linear limit.  We showed that below some small, but not negligible, energy threshold all initial conditions lead to regular motion (at least up to the considered final integration times of $10^9$ time units). This means that below that energy value the multidimensional phase space of the DKG model is covered almost completely by invariant tori and the portion of chaotic regions is  practically negligible, if at all existing. In addition, there exists a higher energy threshold above which all considered initial conditions and system arrangements lead to a chaotic wave-packet spreading. Both these energy thresholds depend on the type of the initial excitation, being higher for single mode excitations. This discrepancy appears because chaos is introduced by the nonlinear interaction of excited NMs, something which happens with greater difficulty
when only one NM is initially excited, as the nonlinearity has to be significantly strong  in order to lead to a considerable interaction between NMs, and consequently to energy spreading.  On the other hand, a single site excitation excites from the beginning of the evolution more than one NMs and thus, a smaller energy is needed to make the interaction of these excited modes significant enough to introduce chaotic dynamics and to permit the delocalization of the wave-packet.

An important outcome of our study was the distinction  of the chaotic cases to two different categories of dynamical behaviors, namely  cases leading to chaotic localization and cases resulting to chaotic spreading of energy, as was done for example in \cite{NTRSA19} for a strongly disordered lattice. We showed that, as we move away from the linear system (when energy grows), chaotic dynamics becomes relevant for the DKG model and localized chaos prevails. In such cases, a few lattice sites are excited, performing a localized chaotic motion without any diffusion of energy to the other degrees of freedom and  oscillators taking place. This is a phenomenon similar to what is called `{\it stable chaos}' in dynamical astronomy \cite{MN92,M93,MF95} where in models of Solar System dynamics it was found that in cases of chaotic motion, different degrees of freedom behave very differently so that diffusion of motion is detected in some of them, while others seem to be insensitive to the dynamics. We also found that the percentage of chaotic orbits exhibiting localized chaos for small $H$ values is larger for single site excitations,  while both single site and single mode excitations lead to spreading chaos as $H$ increases. 

A main advantage of the GALI$_2$ method over the computation of the ftmLCE \eqref{eq:ftMLE} is its ability to identify chaos much more clearly. This happens because when chaotic behavior appears the index tends exponentially fast to zero (something which can be  easily identified numerically),  while the ftmLCE starts showing signs of deviation from the power law $\Lambda_1(t) \propto t^{-1}$, which characterizes regular motion. This deviation is more difficult to be recognized as typically, it would require the inspection of the time evolution of the index, something which is not needed for GALI$_2$ as the computation of its current numerical value is sufficient to identify chaos, without any doubt.

Exploiting further the properties of the GALI$_2$ method we also demonstrated its ability to define a characteristic chaoticity timescale, for the cases of the weak and strong chaos spreading regimes, for which previous studies of the system's ftmLCE indicated a slowing down of chaotic dynamics \cite{SGF13,SMS18}. More specifically, registering the duration $T_G$ of the time intervals needed for the GALI$_2$ to become practically zero (actually $\leq 10^{-8}$), after successive reinitializations of its value to GALI$_2=1$ through the  introduction of  two orthonormal deviation vectors for its computation,  we found that $T_G$ constantly grows, showing in this way the slowing down of chaotic dynamics. The $T_G$ timescale characterizes the strength of the chaotic process in the DKG model, and its increase is very well described by the power laws $T_G \propto t^{0.25}$ and $ T_G \propto t^{0.3}$, for, respectively, the weak and strong chaos cases, showing exactly the same behavior with the so-called Lyapunov time \eqref{eq:L_t_DKG} \cite{SGF13,SMS18}. Thus, GALI$_2$ can be also used in a computationally efficient way to  describe the chaotic nature of motion of multidimensional systems, as well as in cases where the strength of chaos changes in time (as  happens  in the weak end strong chaos regimes considered here, or in time dependent systems \cite{MBS13}). As a final remark let us note that, although we used the GALI$_2$ method for studying the 1D DKG model, we expect that the implementation of the index also in the extension of the model to two special dimensions will be able to efficiently capture the slowing down of chaos (and consequently the increase of the characteristic chaoticity time scale) observed in \cite{MSS20}.

\section*{Acknowledgements}

We thank the High Performance Computing facility of the University of Cape Town, as well as the Centre for High Performance Computing (CHPC) of South Africa for providing the computational resources needed for obtaining the numerical results of this work. We also thank the two anonymous referees for their constructive criticism, which helped us improve the content and the presentation of our work.

\bibliography{11_SS_KG_GALI.bib}

\end{document}